\documentclass[fleqn,twoside]{article}
\usepackage[headings]{espcrc2}

\usepackage{graphicx}

\newcommand{\AmS}{{\protect\the\textfont2
  A\kern-.1667em\lower.5ex\hbox{M}\kern-.125emS}}

\title{ParFORM: recent development \thanks{Supported by SFB-TR9}}

\author{ 
         M.~Tentyukov\address[KU]{
                           Institut f\"ur Theoretische Teilchenphysik,
                           Universit\"at Karlsruhe, Germany
              }\thanks{On leave from BLTP, Joint Institute
              for Nuclear Research, Dubna, Russia},
        J.A.M.~Vermaseren\address[JV]{NIKHEF, Amsterdam} and
       H.M.~Staudenmaier\addressmark[KU]\address{
                             Interfak. Institut f\"ur Anwendungen der Informatik,
                             Universit\"at Karlsruhe, Germany},
       }

\begin{document}

\begin{abstract}
We report on the status of our project of parallelization of the
symbolic manipulation program FORM.  We have now parallel versions of
FORM running on Cluster- or SMP-architectures. These versions can be
used to run arbitrary FORM programs in parallel.
\vspace{1pc}
\end{abstract}

\maketitle

\section{General conceptions of current version}

FORM \cite{form} is a program for symbolic manipulation of algebraic
expressions specialized to handle very large expressions of millions of
terms in an efficient and reliable way. That is why it is widely used in
Quantum Field Theory, where the calculation of the order of several
hundred (sometimes thousands) of Feynman diagrams is required.

In context with this goal an improvement in computing efficiency is
 very important.  Parallelization is one of the most efficient way to
 increase performance.  So the idea to parallelize FORM is quite
 natural.

ParFORM is the parallel version of FORM developed in Karlsruhe since
1998 \cite{Fliegner}. At present, a number of real
physical applications exist which were performed with the help of ParFORM
\cite{PhysAapps}.

There are some internal mechanisms of FORM that
makes FORM very well suited for parallelization \cite{Fliegner,CASC04}.
The concept of parallelization is
indicated in Fig. \ref{parexample}:
\begin{figure}[ht]
\begin{center}
\includegraphics[width=\linewidth]{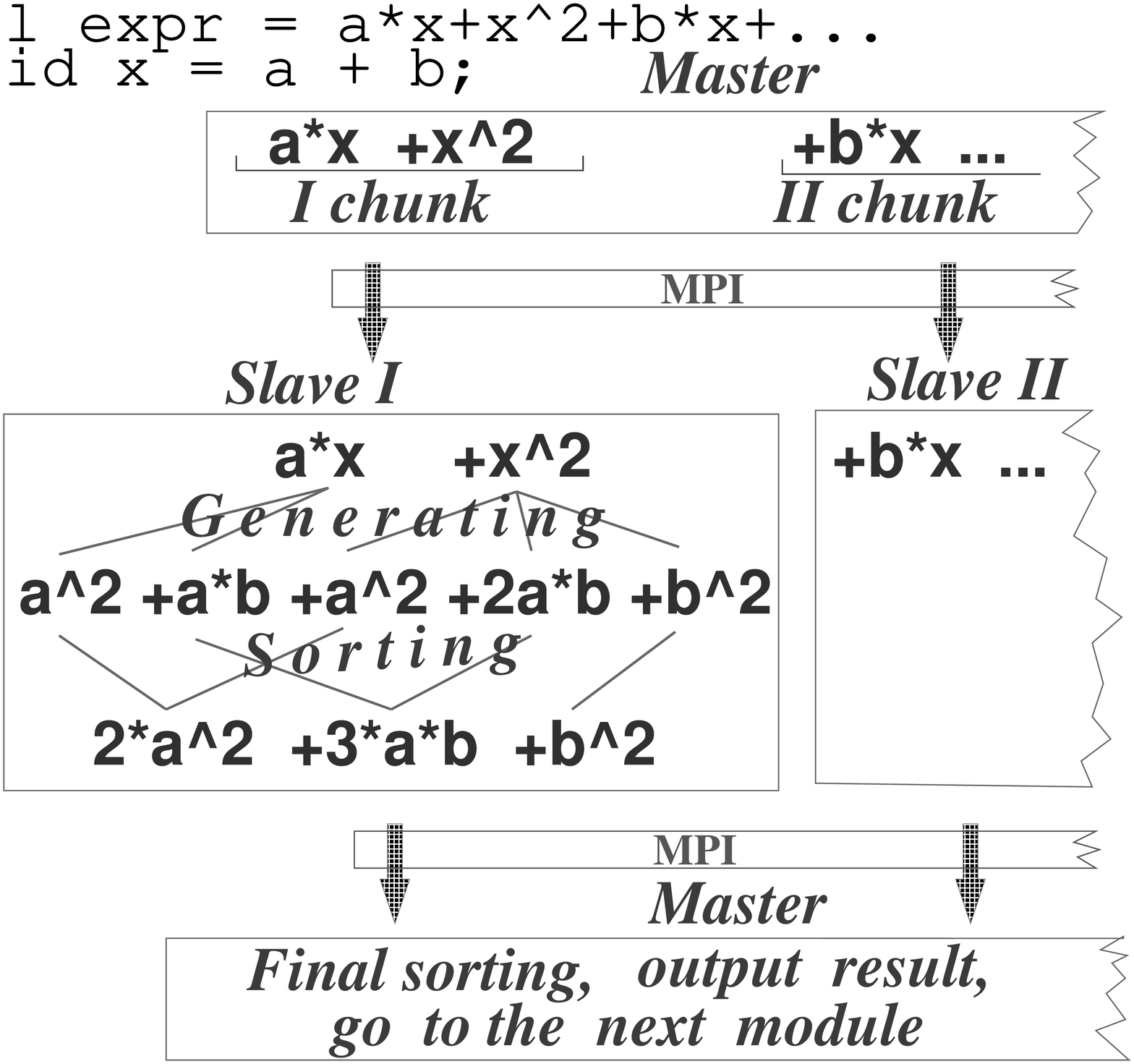}
\end{center}
\caption{
\label{parexample}
General conception of ParFORM: a Master-Slave structure for
parallelization.
\vspace*{-3mm}
}
\end{figure}
Distribute the input terms among available processors, let each of them
perform local operations on its input terms, generate and sort the
arising output terms.
At the end of a module the sorted streams of terms from all processors have
to be  merged to one final output stream again.

\setcounter{footnote}{0}
A {\em master} process initializes the distribution of terms and
finally collects the results The real and time consuming calculations
however are performed by {\em slaves}.
Each process is an independent stream of commands 
operating on independent data.
The master communicates with slaves by means of different message passing
libraries and we use
MPI\footnote{see http://www-unix.mcs.anl.gov/mpi/standard.html}. 

The master simply distributes and collects data, i.e with a lower number
of processors, the master becomes almost idle. For that case one can
try to force the master to participate in real calculations, too. On
the other hand, with an increasing number of slaves, the master spends more
and more time to control slaves, which may lead to early speedup
saturation. Our estimations show that for more than four processors
our Master-Slave model is adequate. 
Since almost all real calculations are performed by slaves
we calculate
speedups normalized to the time spent by program running on two
processors, one master and one slave.

Using the message passing library permits to parallelize FORM on
computer architectures, i.e. with shared (SMP\footnote{Symmetric
MultiProcessor}) and distributed (clusters) memories.

\begin{figure}[ht]
\begin{center}
\includegraphics[width=0.9\linewidth]{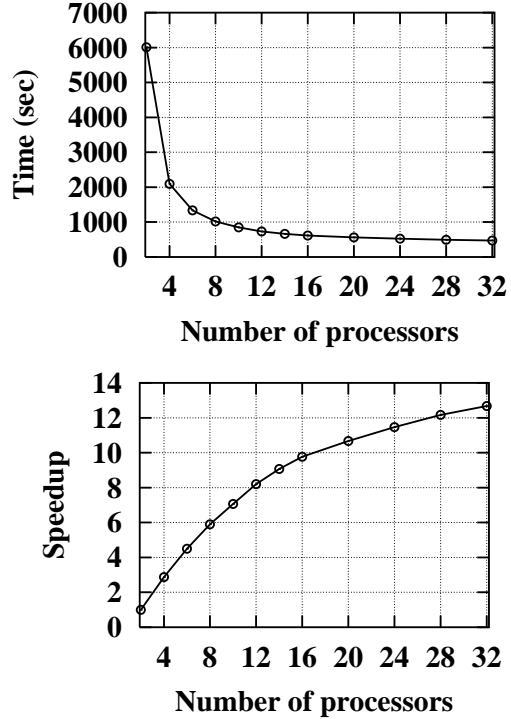}
\end{center}
\caption{
\label{sgi}
Computing time and speedup for the test program BAICER on the
SGI Altix 3700 server with 32x Itanium2 processors (1.3 GHz), a
SMP-type  machine.
\vspace*{-3mm}
}
\end{figure}

The results for the program BAICER\footnote{All benchmarks mentioned
in the paper were made by running the same ``standard'' test example
\cite{CASC04} obtained from a package BAICER developed by P.~Baikov
following methods described in \cite{baikov}.} running on the SMP 
SGI Altix 3700 Server 32x 1.3 GHz/3MB-SC Itanium2 CPUs 
are shown in Fig. \ref{sgi}.
The speedup is almost linear up to 12 processors.
Afterwards the speedup becomes nonlinear but is still
considerable.

\begin{figure}[ht]
\begin{center}
\includegraphics[width=0.9\linewidth]{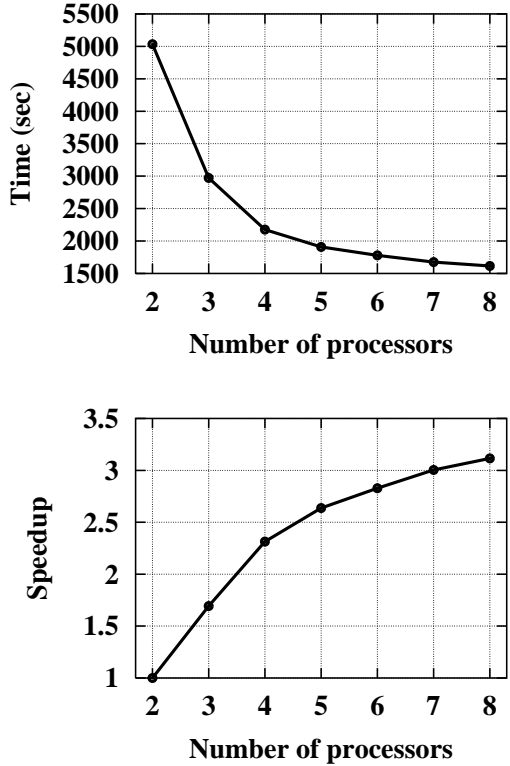}
\end{center}
\caption{
\label{iwarp}
Computing time and speedup for the test program BAICER on the
cluster of dual Intel Xeon, 2.4 GHz, 4x-InfiniBand.
\vspace*{-3mm}
}
\end{figure}

The second architecture is a cluster.
 The results of running
BAICER on an IWR\footnote{The Institute for Scientific Computing of the
Forschungszentrum Karlsruhe.} 
Xeon cluster \cite{iwarpcluster} are shown in Fig. \ref{iwarp}.

The speedup curve has a ``positive'' slope even for more than 8 processors,
but the absolute value of this slope is rather small, and so the speedup is
reasonable only for a few nodes. In case of cluster computers as
\cite{iwarpcluster} it could be better to involve the master processor
in real calculations too, but this should be studied in detail.

\section{ParFORM on SMP}
\label{PFonSMP}
The main disadvantage of the message passing approach is a considerable
overhead due to huge data transfers. On SMP
 computers one can attempt
to get rid of this overhead using e.g. threads \cite{tannenbaum}. 
But in this concept there are several
points which have to be taken into account:

Within a typical SMP machine, all the memory is uniformly available to each
processor, so-called ``Uniform Memory Access''. All memory
accesses are made  by the same shared memory bus. This works quite well for a
relatively small number of CPUs. Increasing the number of CPUs, a
problem with the shared bus
appears due to the  collision rate between
multiple CPU requests on the single memory bus. 

In order to avoid these scalability limits of  SMP architectures,
the ``Non-Uniform Memory Access'' (NUMA) architecture was designed.
NUMA assumes that each processor has its own local memory 
but it can also access memory owned by other processors.

As a result the concept of {\em Memory Affinity} has to be introduced:
memory may be situated at different ``distance'' from the processor.
On a SGI Altix, the ratio of remote to local memory access times varies 
from 1.9 to 3.5, depending on the relative locations of the processor
and the memory.

Usually this is not a problem since nearly all CPU architectures
use a cache to exploit locality of reference in memory accesses.
Because nearly everything is in a cache, often one may safely 
ignore problems resulting from the difference in memory affinity.
But not in the case of FORM as discussed below.

As consequence of the frequent
 ``cache-use'' - as just described - the new
problem of {\em cache coherency} arises: if one of the processors
modifies some piece of data (i.e., performs a ``write'' operation),
then the other processors have access only to an out-of-date copy of
these 
data
stored in their cache. Normally the cache data have to be {\em
invalidated}. Usually, NUMA computers use special-purpose hardware to
maintain cache coherence.  Such systems are called ``cache-coherent
NUMA'', or ccNUMA \cite{ccNUMA}. The worst case for such an approach is
mutual cache invalidation, when two (ore more) processors are writing
to the same memory region.

Let us consider now how ccNUMA could be used for a good
parallelizable problem as the multiplication of two matrices.

Let us take the simplest algorithm is well suited for multithreaded process:
each thread reads one row of the first matrix, one column of the
second matrix and sums up the result of the multiplication in a local-
or even register-variable.
In this case the only instruction is ``read from memory''.
Practically  all arithmetic operations are performed on the local memory, and only
at the end the result is written into the global (shared) memory.

\begin{figure}[ht]
\begin{center}
\includegraphics[width=0.8\linewidth]{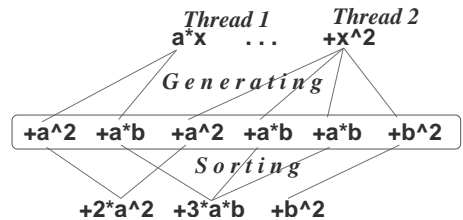}
\end{center}
\caption{
\label{parformcc}
Possible multithreaded approach of FORM parallelization.
\vspace*{-3mm}
}
\end{figure}

Unfortunately, the structure of FORM is quite different. Indeed, let us
suppose that each thread treats one term,  Fig.~\ref{parformcc}.
Then the thread produces a lot of new terms which should be stored in the
shared memory. This would lead to permanent cache invalidation. It indicates
that the internal FORM structure is not well suited for multi-threaded
parallelization on ccNUMA architecture.

\begin{figure}[ht]
\begin{center}
\includegraphics[width=0.7\linewidth]{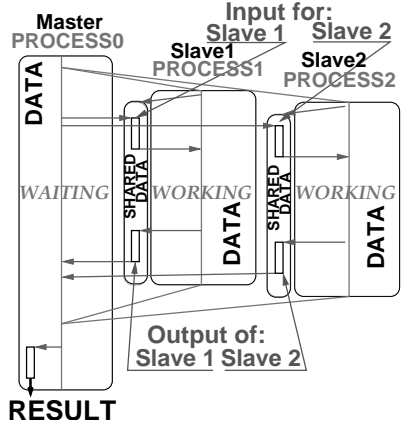}
\end{center}
\caption{
\label{forksM}
Master-Slave approach on SMP without MPI.
\vspace*{-3mm}
}
\end{figure}

Alternatively, one can exploit the multi-process Master-Slave 
structure, 

Instead we could try to use the Master-Slave model discussed above
with multiple processes, but now of course without MPI. 
For communication between the master and slaves we could use shared
memory,
allocating the
shared memory buffers ``close'' to each slave, Fig.~\ref{forksM}.
We refer to this model as ``Shared Memory'' (SM). 
As before, the master splits data into chunks and distributes them
among slaves placing data to shared memory buffers. Slaves
manipulate these data in their local memory, and the (pre-sorted) results
are collected by the master. Here we have explicit control on 
memory affinity, and no message passing bottleneck anymore.

The Master-Slave model permits to optimize the communication between
slaves and the master. For example, no direct communication between slaves
is allowed\footnote{ MPI has ``peer-to-peer'' structure, which is
  much
more complicated.}, so
a lot of optimization available at low level provided the structure is 
restricted by this communication topology.

\section{First results}

We implemented the ideas described in the previous section in ParFORM
(we call it ParFORM-SM) and want to present first results in the
following.

\begin{figure}[ht]
\begin{center}
\includegraphics[width=0.9\linewidth]{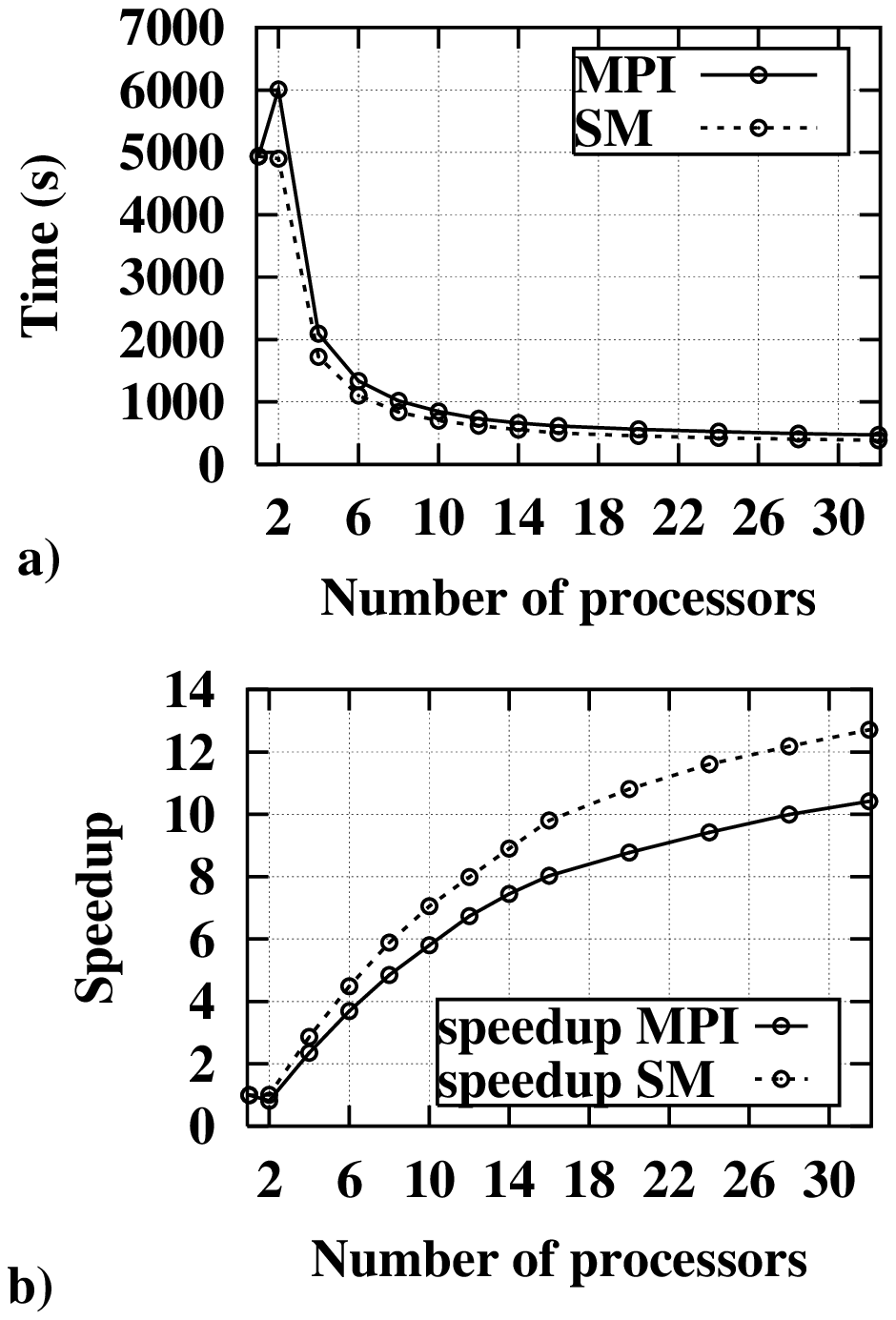}
\end{center}
\caption{
\label{mpi_sm}
Results of running the test program on SGI Altix 3700 with MPI-based
communications (MPI) and with Shared Memory segments  (SM) 
normalized to the sequential version.
\vspace*{-3mm}
}
\end{figure}

In Fig.~\ref{mpi_sm} one can see the comparison with the results from
Fig.~\ref{sgi}.  We can immediately see about 20\% performance
improvement compared with the previous MPI version. But the most important
observation is that now the communication overhead is almost
negligible as can be seen comparing the first two data points in
Fig.~\ref{mpi_sm} a).

Here we normalize the results not to the two-processor time, but to
the time spent by the corresponding sequential version of the
program. Looking at the difference in times between one processor
(sequential program) and two processors for MPI variant (solid line),
we may see about 20\% of performance reduction. The reason is due to
the communication overhead. Indeed, in a two-processor mode the single
slave is doing almost all the job (except the final sorting) and the
program spends some extra time for communication between the master
and the slave.

For the shared memory based program, the difference in one- and 
two-processor regimes is unobservable. This indicates that the
communication overhead has no real significance in this SM model.

Increasing the number of processors, an other bottleneck arises: the time
for final sorting becomes more and more essential. Since this
sorting is performed only by the master, all the slaves are idle during
this stage. This explains the speedup saturation around 30 nodes
both for MPI and SM approaches.

\section{Outlook}

We shortly want to discuss the various aspects of the models and
architectures described before. 

On ccNUMA computers, instead of MPI, we should use the multiprocessed
model (see Sect.~\ref{PFonSMP}) with multiple shared memory segments.
The corresponding shared memory approach was developed, tested and
demonstrates stable performance improvement around 20\%. The
communication overhead is negligible and the main bottleneck is the
final sorting stage, so it seems to be reasonable to parallelize in
future the final sorting process first.

On clusters, there are no alternatives to MPI at the moment. 

In the present cluster version of ParFORM the communication overhead
is quite big and thus it can only be used for a relatively small
number of nodes. In particular, it seems that it would be advantageous
if also the master participates in real calculations.

Colleagues who are interested to use ParFORM software should contact
M.Tentyukov

\end{document}